\documentclass{article}

\usepackage{natbib}

\usepackage{hyperref}
\usepackage{amsmath} % Package for AMS math extensions.

\usepackage{url}

\usepackage{amsmath}
\usepackage{graphicx}
\usepackage{tikz}

\usetikzlibrary{automata,calc,shapes.geometric,positioning,trees,fit,chains,matrix,scopes,arrows,petri}

\tikzset{
	every picture/.style={
		auto,
		>=latex,
		thick,
		bend angle=10,
		label distance=-1pt,
		initial text={},
	},
	every state/.style={
		inner sep=3pt,
		minimum size=.4cm,
	},
	every loop/.style={
		looseness=6
	},
	transitionsystem/.style={automaton},
	automaton/.style={
		->,
		every state/.style={ellipse},
	}
}

\newenvironment{mytabbing}{\begin{tabbing}\hspace*{\leftskip}\=\+\kill}{\end{tabbing}}

\def\sep{[\!]}

\newcommand{\bfe}[1]{\begin{bfseries}\emph{#1}\end{bfseries}}

\newcommand{\fa}{\mbox{$\forall$}}

\newcommand{\TERM}      {\mbox{$T\!E\!R\!M$}}
\newcommand{\al}{{\mbox{$\alpha$}}}
\newcommand{\Eff}       {\mbox{$\mathit{Eff}$}}
\newcommand{\ES}{\mbox{$\emptyset$}}

\newcommand{\po}{\mbox{$\ \sqsubseteq\ $}}

\newcommand{\bx}{\mbox{$\bar{x}$}}
\newcommand{\by}{\mbox{$\bar{y}$}}

\newcommand{\IF}{\mbox{{\bf if}\ }}
\newcommand{\FI}{\mbox{{\bf fi}}}
\newcommand{\DO}{\mbox{{\bf do}\ }}
\newcommand{\OD}{\mbox{{\bf od}}}
\newcommand{\IFP}{\mbox{$\IF B_1 \ra S_1 \sep \LL \sep B_n \ra S_n\ \FI$}}
\newcommand{\DOP}{\mbox{$\DO B_1 \ra S_1 \sep \LL \sep B_n \ra S_n\ \OD$}}

\newcommand{\IFPa}{\mbox{$\IF \sep^n_{i=1}\ B_i \ra S_i\ \FI$}}
\newcommand{\DOPa}{\mbox{$\DO \sep^n_{i=1}\ B_i \ra S_i\ \OD$}}

\newcommand{\WHILE}{\mbox{{\bf while}\ }}

\newcommand{\THEN}{\mbox{\ {\bf then}\ }}
\newcommand{\ELSE}{\mbox{\ {\bf else}\ }}
\newcommand{\FOR}{\mbox{{\bf for}\ }}
\newcommand{\TO}{\mbox{{\bf to}\ }}

\newcommand{\tra}{\mbox{$\:\rightarrow^*\:$}}

\newcommand{\ra}{\mbox{$\:\rightarrow\:$}}
\newcommand{\A}{\mbox{$\ \wedge\ $}}
\newcommand{\LL}{\mbox{$\ldots$}}
\newcommand{\LLn}{\mbox{$1,\ldots,n$}}

\newcommand{\ITE}[3]{\mbox{$\IF {#1} \THEN {#2} \ELSE {#3}\ \FI$}}
\newcommand{\IT}[2]{\mbox{$\IF {#1} \THEN {#2}\ \FI$}}
\newcommand{\WD}[1]{\mbox{$\WHILE {#1}\ \DO$}}

\newcommand{\C}[1]{\mbox{$\{{#1}\}$}}

\newcommand{\SENDER}        {\mbox{$S\!E\!N\!D\!E\!R$}}
\newcommand{\RECEIVER}        {\mbox{$R\!EC\!EIV\!E\!R$}}
\newcommand{\FILTER}        {\mbox{$F\!I\!LT\!E\!R$}}

\newcommand{\BLANK}        {\mbox{`\ '}}
\newcommand{\AST}        {\mbox{`$*$'}}

\newcommand{\NI}{\noindent}

\newcommand{\PP}{\mbox{$[S_1 \| \LL \| S_n]$}}

\begin{document}

\title{Nondeterminism and Guarded Commands\footnote{This article appeared as Chapter \cite{AO22} in the book \cite{AH22}. }}

\author{Krzysztof R. Apt\footnote{CWI, Amsterdam, The Netherlands and MIMUW, University of Warsaw, Warsaw, Poland} \\
Ernst-R\"{u}diger Olderog\footnote{University of Oldenburg, Oldenburg, Germany}
}
\date{}

\maketitle

\section{Introduction}

The purpose of this chapter is to review Dijkstra's contribution to
nondeterminism by discussing the relevance and impact of his guarded
commands language proposal. To properly appreciate it we
explain first the role of nondeterminism in computer science at the
time his original article \cite{Dij75} appeared.

The notion of computability is central to computer science. It was
studied first in mathematical logic in the thirties of the last
century. Several formalisms that aimed at capturing this notion were
then proposed and proved equivalent in their expressive power:
$\mu$-recursive functions, lambda calculus, and Turing machines, to
mention the main ones.

Alan Turing alluded to nondeterminism in his
original article on his machines, writing

\begin{quote}
  For some purposes we might use machines (choice machines or
  $c$-machines) whose motion is only partially determined by the
  configuration. [$\dots$] When such a machine reaches one of these
  ambiguous configurations, it cannot go on until some arbitrary
  choice has been made by an external operator.\cite[page 232]{Tur37}
\end{quote}

However, subsequently he limited his exposition to deterministic machines
and it seems that the above option has not been pursued for quite some time.
As pointed out in \cite{AB09}, an interesting account of nondeterminism,
in the classic book by Martin
Davis, Turing machines were used with a restriction that
\begin{quote}
  no Turing machine will ever be confronted with two different
  instructions at the same time \cite[page 5]{Dav58}.
\end{quote}

Another early book by Hermes [\citeyear{Her65}] introduced the theory of
computability with deterministic Turing machines as equivalent to
the $\mu$-recursive functions.  Further, as pointed out in \cite{STB89},
another helpful study of nondeterminism,
in a standard comprehensive introduction to the recursion theory by
Hartley Rogers Jr., Turing machines were explicitly assumed to be
deterministic:
\begin{quote}
[$\dots$] Finally, the device is to be constructed that it behaves according to 
 a finite list of deterministic rules [$\dots]$ \cite[page 13]{Rog87}
\end{quote}

It seems that a systematic study of addition of nondeterminism to formalisms
concerned with computability is due to computer scientists.  What
follows is a short exposition of such formalisms. Then we discuss
Dijkstra's contribution and its relevance. We conclude by providing a
brief account of other approaches to nondeterminism that followed.

Literature on nondeterminism in computer science is really extensive.
The word `nondeterminism' yields 421 hits in the DBLP database, while
`nondeterministic' results in 1141 matches.  Our intention was not to
provide a survey of the subject but rather to sketch a background
against which one can adequately assess Dijkstra's contribution to the subject.

When working on our book \cite{ABO09}, written jointly with Frank de
Boer, we found that guarded commands form a natural `glue' that
allowed us to connect the chapters on parallel programs, distributed
programs, and fairness into a coherent whole.  This explains the regular
references to this book in the second half of this chapter.

\section{Avoiding nondeterminism}

We begin with two early formalisms in which nondeterminism
is present but the objective is to use them in such a way that
it is not visible in the outcome.

\subsection*{Grammars}

\begin{figure}[ht]

\begin{center} 
\setlength{\unitlength}{1mm} 
\begin{picture}(90,35) 
\put(10,30){\makebox(5,5){$S$}}

  \put(5,25){\line(1,1){5}} \put(12.5,25){\line(0,1){5}} \put(20,25){\line(-1,1){5}}

  \put(0,20){\makebox(5,5){$S$}} \put(10,20){\makebox(5,5){+}} \put(20,20){\makebox(5,5){$S$}}

  \put(2.5,15){\line(0,1){5}} \put(15,15){\line(1,1){5}} \put(22.5,15){\line(0,1){5}} \put(30,15){\line(-1,1){5}}

  \put(0,10){\makebox(5,5){$x$}} \put(10,10){\makebox(5,5){$S$}} \put(20,10){\makebox(5,5){$*$}}
  \put(30,10){\makebox(5,5){$S$}}

  \put(12.5,5){\line(0,1){5}} \put(32.5,5){\line(0,1){5}}

  \put(10,0){\makebox(5,5){$y$}} \put(30,0){\makebox(5,5){$z$}}

  \put(35,20){\makebox(20,5){and}}

  \put(75,30){\makebox(5,5){$S$}}

  \put(70,25){\line(1,1){5}} \put(77.5,25){\line(0,1){5}} \put(85,25){\line(-1,1){5}}

  \put(65,20){\makebox(5,5){$S$}} \put(75,20){\makebox(5,5){$*$}} \put(85,20){\makebox(5,5){$S$}}

  \put(57.5,15){\line(1,1){5}} \put(67.5,15){\line(0,1){5}} \put(77.5,15){\line(-1,1){5}} \put(87.5,15){\line(0,1){5}}

  \put(55,10){\makebox(5,5){$S$}} \put(65,10){\makebox(5,5){+}} \put(75,10){\makebox(5,5){$S$}}
  \put(85,10){\makebox(5,5){$z$}}

  \put(57.5,5){\line(0,1){5}} \put(77.5,5){\line(0,1){5}}

  \put(55,0){\makebox(5,5){$x$}} \put(75,0){\makebox(5,5){$y$}} 
\end{picture} 
\end{center}
\caption{\label{fig:Ambiguity} Two different parsing trees for $x+y*z$
using grammar (\ref{eq:ambig-grammar}).}
\end{figure}
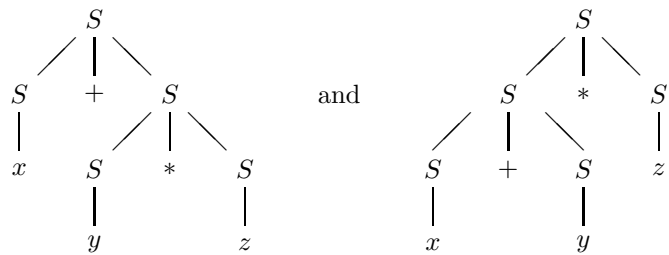

Mechanisms that are nondeterministic from the very start are
\emph{grammars} in the Chomsky hierarchy, \cite{Cho59}. 
For example, the set of arithmetic expressions with 
variables $x,y,z$, operators $+$ and $*$, and brackets $($ and $)$ 
as terminal symbols can be defined by the following context-free grammar 
using the start symbol $S$ as its only nonterminal:
\begin{equation} \label{eq:ambig-grammar}
 S ::= x\ |\ y\ |\ z\ |\ S+S\ |\ S*S\ |\ (S).
\end{equation}
Here it is natural to postulate
that in a derivation step $\vdash$, the nonterminal
$S$ can be replaced by any of the above right-hand sides.
In particular, the arithmetic expression $x+y*z$ has two different 
\emph{leftmost derivations},
where always the leftmost occurrence of $S$ is replaced:
\begin{eqnarray} 
  \label{eq:derivation-1}
   S \vdash S+S \vdash x+S \vdash x+S*S \vdash x+y*S \vdash x+y*z \\
  \label{eq:derivation-2}
   S \vdash S*S \vdash S+S*S \vdash x+S*S \vdash x+y*S \vdash x+y*z
\end{eqnarray}
In a compiler, derivation (\ref{eq:derivation-1}) corresponds to a
parsing tree shown on the left-hand side of Figure~\ref{fig:Ambiguity}, 
giving priority to the operator $*$, 
while derivation (\ref{eq:derivation-2}) corresponds to a parsing tree on the right-hand side,
giving priority to $+$.

Thus by definition, the above grammar is \emph{ambiguous}.  When using
context-free grammars to define syntax of a programming language, one
is interested in \emph{unambiguous} grammars, meaning that each word
(here: program) has only one parsing tree. 
So in grammars, nondeterminism (in the application of the production
rules) is allowed, but the objective is that it does not lead to
ambiguities in the above sense.

\begin{figure}[ht]

\begin{center} 
\setlength{\unitlength}{1mm} 
\begin{picture}(35,45) \put(10,40){\makebox(5,5){$E$}}

  \put(5,35){\line(1,1){5}} \put(12.5,35){\line(0,1){5}} \put(20,35){\line(-1,1){5}}

  \put(0,30){\makebox(5,5){$E$}} \put(10,30){\makebox(5,5){+}} \put(20,30){\makebox(5,5){$T$}}

  \put(2.5,25){\line(0,1){5}} \put(15,25){\line(1,1){5}} \put(22.5,25){\line(0,1){5}} \put(30,25){\line(-1,1){5}}

  \put(0,20){\makebox(5,5){$T$}} \put(10,20){\makebox(5,5){$T$}} \put(20,20){\makebox(5,5){$*$}}
  \put(30,20){\makebox(5,5){$F$}}

  \put(2.5,15){\line(0,1){5}} \put(12.5,15){\line(0,1){5}} \put(32.5,15){\line(0,1){5}}

  \put(0,10){\makebox(5,5){$F$}} \put(10,10){\makebox(5,5){$F$}} \put(30,10){\makebox(5,5){$z$}}

  \put(2.5,5){\line(0,1){5}} \put(12.5,5){\line(0,1){5}}

  \put(0,0){\makebox(5,5){$x$}} \put(10,0){\makebox(5,5){$y$}} 
\end{picture} 
\end{center}
\caption{\label{fig:UniqueParsing} Unique parsing tree for $x+y*z$
using grammar (\ref{eq:unambig-grammar}).}

\end{figure}
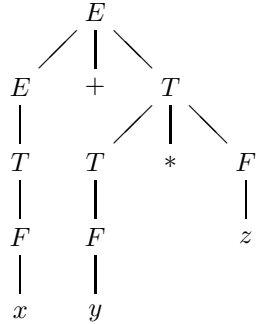

A context-free grammar that allows nondeterminism but is unambiguous 
uses three nonterminals, $E$ (for `expression'), $T$ (for `term'), and
$F$ (for `factor'), where $E$ is the start symbol, 
and the following production rules:
\begin{eqnarray} \label{eq:unambig-grammar}
  E & ::= & T\ |\  E +T \\ 
  \notag
  T & ::= & F\ |\ T*F \\ 
  \notag
  F & ::= & (E)\ |\ x\ |\ y\ |\ z 
\end{eqnarray}
This grammar generates the same set of arithmetic expressions as the one above,
but is unambiguous. In particular, the arithmetic expression $x + y * z$ has now 
only one leftmost derivation corresponding to the unique parsing tree shown in 
Figure~\ref{fig:UniqueParsing}.
The grammar encodes the fact that the operator $*$ has a higher priority than $+$
and that expressions with the same operator are evaluated from left to right.
It can be generalized to a pattern dealing with
any set of infix operators with arbitrary priority among them.

\subsection*{Abstract reduction systems}

Another simple formalism that allows nondeterminism are abstract
reduction systems.  Formally, an \emph{abstract reduction system} is a
pair $(A,\ra)$ where $A$ is a set and $\ra$ is a binary relation on
$A$.  If $a \ra b$ holds, we say that $a$ can be \emph{replaced} by
$b$.  In this setting nondeterminism means that an element can be
replaced in various ways.

There are several important examples of abstract reduction systems, in
particular \emph{term rewriting systems}, with \emph{combinatory
  logic} and \emph{$\lambda$-calculus}, and some functional languages
as best known examples (see, e.g., \cite{Ter03}).

Let $\tra$ denote the reflexive transitive closure of $\ra$.  An
element $a \in A$ is said to be in \emph{normal form} if for no
$b \in A$, $a \ra b$ holds. If $a \tra b$ and $b$ is in normal form,
then $b$ can be viewed as a \emph{value} of $a$ obtained by means of
an abstract computation consisting of a repeated application of the
$\ra$ relation.  In general one is interested in abstract reduction
systems in which each element has at most one normal form, so that the
notion of a value can be unambiguously defined.
One says then that the system has the \emph{unique normal form} (UN, in short).

To establish UN it suffices to establish the \emph{Church-Rosser property} (CR, in short).
It states that for all $a,b,c \in A$

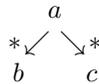
\begin{center}
$a$                                             \\
$*\!\swarrow$ $\searrow\! *$                    \\
$b$\ \ \ \ \ \ \ $c$
\end{center}

\NI
implies that for some $d \in A$

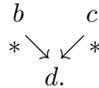
\begin{center}
$b$\ \ \ \ \ \ \ $c$                            \\
$*\!\searrow$ $\swarrow\! *$                    \\
$d$.
\end{center}
Indeed, CR implies UN.

Several important term rewriting systems, including combinatory logic and
$\lambda$-calculus, have CR, and hence UN, see, e.g., \cite{Ter03}.

In this area the interest in UN means that one is interested in
deterministic outcomes \emph{in spite of} the nondeterminism that is
present, that is, one aims at showing that the nondeterminism is
\emph{inessential}.

\section{Angelic nondeterminism}

We now proceed with a discussion of formalisms in which addition of
nondeterminism allowed one to extend their expressiveness. These
formalisms share a characteristic that one identifies
successes and failures and only the former count.  This kind of
nondeterminism was later termed \emph{angelic nondeterminism}.

\subsection*{Nondeterministic finite automata and Turing machines}

The first definitions of finite-state automata
required deterministic transition functions, as noted in
\cite{HopcroftU79}. It took the insight of Rabin and Scott to
introduce \emph{nondeterministic} finite automata in their seminal
paper \cite{RS59}. Such an automaton has choices in its moves: at each
transition it may select one of several possible next states.  They
motivated their definition as follows: ``The main advantage of these
machines is the small number of internal states that they require in
many cases and the ease in which specific machines can be described.''
They proved by their famous \emph{power set construction} (that uses
as the set of states the powerset of the original set of states) that
for each nondeterministic automaton a deterministic one can be
constructed that accepts the same set of finite words, however the
deterministic one may have exponentially more states.  Crucial is here
their definition of acceptance: a nondeterministic automaton
\emph{accepts} a word if there is \emph{some} successful run of the automaton
from an initial to a final state, processing the word symbol by
symbol. Thus only a success counts, while failures do not matter.

This kind of nondeterminism has also been introduced for other types
of machines, in particular pushdown automata and Turing machines,
leading to the definition of various complexity classes discussed
at the end of this section.

\subsection*{McCarthy's ambiguity operator}

Probably the first proposal to add nondeterminism to a programming
language is due to John McCarthy who introduced in \cite{McC63} an
`ambiguity operator' $amb(x,y)$ that, given two expressions $x$ and
$y$, nondeterministically returns the value of $x$ or of $y$ when both
are defined, and otherwise whichever is defined.  (McCarthy did not
explain what happens when both $x$ and $y$ are undefined, but the most
natural assumption is that $amb(x,y)$ is then undefined, as
well.) In particular, $amb(1,2)$ yields 1 or 2.

McCarthy was concerned with the development of a functional language,
so in his formulation programs were expressions, possibly defined by
recursion. As an example of a program that uses the ambiguity operator
he introduced the function $less(n)$ that assigns to each natural
number $n$ any nonnegative integer less than $n$.  The function was
defined recursively by:
\[
less(n) = amb(n-1, less(n-1)).
\]

McCarthy did not discuss this function in detail, but note that its
definition involves a subtlety because of its use of undefined
values.  For the smallest value in its domain, namely~1, we get
$less(1) = amb(0, less(0)) = 0$, since by definition $less(0)$ is
undefined. For the next value, so 2, we get
$less(2) = amb(1, less(1)) = amb(1, 0)$, which yields 0 or 1.  Next,
$less(3) = amb(2, less(2)) = amb(2, amb(1,0))$, which yields 0 or 1 or
2, etc.
This may be the first example of a program that uses nondeterminism.

McCarthy's used his ambiguity operator to extend computable functions
by nondeterminism to what he called \emph{computably ambiguous
  functions}. His work soon inspired first
proposals of systems and programming languages that incorporated some
form of nondeterminism, mainly to express concisely search problems, see,
e.g., \cite{SE73} for an account of some of them.

Sometime later, the authors of \cite{ZMC87} extended a dialect of Lisp
with McCarthy's nondeterministic operator denoted by \texttt{AMB}.
The addition of nondeterminism was coupled with a dependency-directed
backtracking, triggered by a special expression \texttt{FAIL} that has
no value.  The resulting language was called SCHEMER.  This addition
of nondeterminism to Lisp was later discussed in the book \cite{AS96}
that used a dialect of Lisp called Scheme.

\subsection*{Floyd's approach to nondeterministic programming}

Another approach to programming that corresponds to the type of
nondeterminism used in the nondeterministic Turing machines was
proposed by Robert Floyd in \cite{Flo67}. He began his article by a
fitting quotation from a famous poem `The Road Not Taken' by Robert
Frost:

\begin{quote}
Two roads diverged in a yellow wood, \\
And sorry I could not travel both \\
And be one traveler, long I stood \\
And looked down one as far as I could \\
To where it bent in the undergrowth;
\end{quote}

Floyd's idea was to add nondeterminism
to the conventional flowcharts  by

\begin{itemize}
\item using a nondeterministic assignment
\[
  \mbox{$x := \emph{choice}(t)$},
\]
that assigns to the variable $x$ an arbitrary positive integer 
of at most the value of the integer expression $t$,

\item labeling all termination points as \emph{success} or \emph{failure}.

\end{itemize}

Thus augmented flowcharts can generate several execution sequences, however,
only those that terminate in a node labeled \emph{success} are considered
as the computations of the presented algorithm.

Using a couple of examples Floyd showed how these two additions can
lead to simple flowchart programs that search for a solution through
exhaustive enumeration and which otherwise would have to be programmed
using backtracking combined with appropriate bookkeeping.

Jacques Cohen illustrated in \cite{Coh79} Floyd's approach using
conventional programs in which one uses the above nondeterministic
assignment statement and a \textbf{fail} statement that corresponds
to a node labeled \emph{failure}. Any computation that reaches the
\textbf{fail} statement terminates improperly (it \emph{aborts}).  The
label \emph{success} is unneeded as it is implicitly modeled by a
terminating computation that does not abort.  We call such
computations \emph{successful}.  In this approach a program is correct
if \emph{some} successful computation establishes the assumed
postcondition.

We illustrate this approach by means of one of Floyd's examples, the
problem of \emph{eight queens} in which one is asked to place 8 queens
on the chessboard so that they do not attack each other.  The program
looks as follows, where $a$, $b$ and $c$ are integer arrays with
appropriate bounds, and initialized for all indices to 0:

\begin{mytabbing}
\qquad\qquad
\=  \FOR $col:=1 $ \TO 8 \DO \\
\> \ \ \= $row := \emph{choice}(8)$; \\
\>     \> $\IT{a[row] = 1 \vee b[row+col] = 1 \vee c[row-col] = 1}{\textbf{fail}}$; \\
\>     \> $a[row] := 1$; \\
\>     \> $b[row+col] := 1$; \\
\>     \> $c[row-col] := 1$ \\
\>  \OD 
\end{mytabbing}
Subscripted variables have the following interpretation:

$a[i] = 1$ means that a queen was placed in the $i$th row,

$b[i] = 1$ means that a queen was placed in the $i$th $\searrow$ diagonal, 

$c[i] = 1$ means that a queen was placed in the $i$th $\swarrow$ diagonal,

\NI
where the $\searrow$ diagonals are the ones for which the sum of
the coordinates is the same and the $\swarrow$ diagonals are the ones
for which the difference of the coordinates is the same.  Upon
successful termination a solution is produced in the form of a
sequence of 8 values that are successively assigned to the variable
$row$; these values correspond to the placements of the queens in the
columns 1 to 8.

Thanks to the nondeterministic assignment and the \textbf{fail}
statement this program can generate several computations, including
ones that abort.  The successful computations generate 
precisely all solutions to the eight queens problem.

Floyd showed that one can convert his augmented flowcharts 
to conventional flowcharts by means of a
generic transformation that boils down to an implementation of
backtracking.  This way one obtains a deterministic algorithm that
generates a solution to the considered problem but it is easy to
modify his transformation so that all solutions are generated.  Of
course, such a transformation can also be defined for the
programs considered here.

\subsection*{Logic programming and Prolog}

In the early seventies angelic nondeterminism was embraced in a novel
approach to programming that combined the idea of \emph{automatic
  theorem proving} based on the use of relations, with built-in
automatic backtracking. It was realized in the programming language
Prolog, conceived and implemented by Alain Colmerauer and his team
(see the historic account in \cite{CR96}), while its theoretical
underpinnings, called \emph{logic programming}, were provided by
Robert Kowalski in \cite{Kow74}.

A detailed discussion of logic programming and Prolog is out of scope
of this chapter, but to illustrate the nondeterminism present in
this approach to programming consider a simple logic programming
program that appends two lists. It is defined by means of two
\emph{clauses}, the first one unconditional and the second one
conditional: \smallskip

\NI
\texttt{append([], Xs, Xs).} \\
\texttt{append([X | Xs], Ys, [X | Zs])} $\leftarrow$  \texttt{append(Xs, Ys, Zs)}. 

\smallskip

Here \texttt{append} is a name of a relation, \texttt{X, Xs, Ys} and
\texttt{Zs} are the variables, \texttt{[]} denotes the empty list, and
\texttt{[X | Xs]} denotes the list with the head \texttt{X} and the
tail \texttt{Xs}.

The first clause states that the result of appending \texttt{[]} and
the list \texttt{Xs} is \texttt{Xs}. The second clause is a reverse
implication stating that if the result of appending the lists
\texttt{Xs} and \texttt{Ys} is \texttt{Zs}, then the result of
appending the lists \texttt{[X | Xs]} and \texttt{Ys} is \texttt{[X |
  Zs]}. 

In general, a program is a set of clauses which are built from
\emph{atomic queries}. In the above program atomic queries are of the
form \texttt{append(s, t, u)}, where \texttt{s, t, u} are expressions
built out of the variables and the constant \texttt{[]} using the list
formation operation \texttt{[.$\mid$.]}.  A \emph{query} is a
conjunction of atomic queries. A program is activated by executing a
query, which is a request to evaluate it w.r.t.~the considered
program.  We do not discuss here the underlying computation model; it
suffices to know that the computation searches for an instance of the
query that logically follows from the program. If such an instance is
found, one says that a query \emph{succeeds} and otherwise that it
\emph{fails}.

In logic programming nondeterminism arises in two ways, by the fact
that relations can be defined using several clauses, and by the choice
of the atomic query to be evaluated first.  At the abstract level this
is the same form of nondeterminism as the one used in Floyd's
approach: a computation succeeds if at all choice points the right
selections are made.

Since a query can succeed in several ways, several
solutions can be generated.  Consider for example the query
\smallskip

\texttt{append(\_, Zs, [mon, tue, wed, thu,  fri]), append(Xs, \_, Zs),}

\smallskip

\NI
in which we use Prolog's convention according to which each
occurrence of `\texttt{\_}' stands for a different (\emph{anonymous},
i.e, irrelevant) variable and the comma is used (between the atomic
queries) instead of the conjunction sign.  Intuitively, this query
stipulates that \texttt{Xs} is a prefix of a suffix of the list
\texttt{[mon, tue, wed, thu, fri]}.  Successful computations of this
query w.r.t.~the above program generate in \texttt{Xs} all possible
sublists of this list.

In Prolog nondeterminism present in the computation model of logic
programming is resolved by stipulating that the first clause and the
first atomic query from the left are selected, and by providing a
built-in automatic backtracking that allows the computation to recover
from a failure.

\subsection*{Does angelic nondeterminism add expressive power?}
\label{subsec:?}

After this discussion of nondeterministic programming let us look at the
idea of angelic nondeterminism through the lens of computing and structural complexity.

We mentioned already in the introduction that nondeterminism was not
considered in the original formalizations of the notion of a
computation.  But once one considers restricted models of
computability, nondeterminism naturally arises.  An early example is
the characterization of formal languages in the Chomsky hierarchy of
formal grammars. It distinguishes four types of grammars that
correspond to four levels of formal languages of increasing
complexity: regular, context-free, context-sensitive, and recursively
enumerable languages, denoted by Type-3, Type-2, Type-1,
and Type-0, respectively.

The need for nondeterminism arises in connection with the
characterization of these classes by means of an automaton.  Whereas
Type-0 languages at the top of this hierarchy are the ones accepted by
the deterministic Turing machines, and Type-3 languages at the bottom,
i.e., regular languages, are the ones accepted by the deterministic finite
automata, the characterization of the remaining two levels calls for the use
of nondeterministic automata or machines.

In particular Type-2 languages, i.e., context-free languages, cannot be
accepted by deterministic pushdown automata.  A pushdown automaton
extends a finite automaton by an unbounded stack or pushdown list that
is manipulated in a `first in -- last out' fashion while scanning a
given input word letter by letter.  Such an automaton is called
\emph{deterministic} if for each control state, each symbol at the top
of the stack, and each input symbol, it has at most one possible move;
it is called \emph{nondeterministic} if more than one move is allowed.

A standard example is the language of all \emph{palindromes} over
letters $a$ and $b$, i.e., words that read the same forward and
backward, like $abba$.  This language is context-free, i.e., it can be
generated by a context-free grammar, but it cannot be accepted by a
deterministic pushdown automaton. The intuitive reason is that while
checking an input word letter by letter, one has to `guess' when the
middle of the word has been reached, so that one can test that from now
on the letters occur in the reverse order, by referring to the constructed
pushdown list. 

However, context-free languages can be characterized by means of
nondeterministic pushdown automata. Such a characterization refers to
angelic nondeterminism because it states that a word is generated by
the language iff it can be accepted by \emph{some} computation of the
automaton.

Similarly, nondeterminism is used to characterize Type-1 languages, i.e.,
context-sensitive languages: they are the ones that can be recognized
by linear bounded nondeterministic Turing machines.

While deterministic and nondeterministic Turing machines accept the
same class of languages, the Type-0 languages, there is a difference
when time complexity is considered.  Probably the most known are the
complexity classes P and NP of problems that can be solved in
polynomial time by deterministic Turing machines and nondeterministic
Turing machines, respectively.
The class NP was introduced by Stephen Cook in \cite{Coo71} and Leonid
Levin in \cite{Lev73}.  The intuition is that a problem is in NP if it
can be solved by first (nondeterministically) guessing a candidate
solution and then checking in polynomial time whether it is indeed a
solution. Following the paradigm of angelic nondeterminism, wrong
guesses do not count. The famous open problem posed by Cook is whether
P $=$ NP holds.

In contrast, when instead of time, space is considered as the
complexity measure, it is known that there is no difference between
the resulting classes PSPACE and NSPACE of problems that can be solved
in polynomial space by deterministic Turing machines and
nondeterministic Turing machines, respectively. This is a consequence
of a result by Savitch [\citeyear{Sav70}].

We conclude that the addition of angelic nondeterminism can, but does not have to,
increase expressive power of the considered model of computability.

\section{Guarded commands}

One of us (KRA) met Edsger Dijkstra for the first time in Spring 1975,
while looking for an academic job in computer science in the
Netherlands. During a meeting at his office at the Technical
University of Eindhoven, Dijkstra handed him a copy of his EWD472
titled \emph{Guarded Commands, Nondeterminacy and Formal Derivation of
Programs}. It appeared later that year as \cite{Dij75}.  This short,
five-pages long, paper introduced two main ideas: a small programming
language, now called \emph{guarded commands}, and its semantics, now
called the \emph{weakest precondition semantics}. Both were new
ideas of great significance.

In this chapter we focus on the guarded commands; Chapters
\cite{Gri22} 
% \ref{chapter:gries} written by David Gries,
and \cite{Hah22} in the book \cite{AH22}
% \ref{chapter:haehnle},
%written by Reiner H\"{a}hnle,
discuss the
weakest precondition semantics.
The essence of guarded commands boils down to two new programming constructs:
\begin{itemize}
\item   \bfe{alternative command}
\[ S::=\IFP, \]
\item \bfe{repetitive command}
\[ S::=\DOP. \]
\end{itemize}

We sometimes abbreviate these commands to
\[ \IFPa\ \mbox{and}\ \DOPa. \]
A Boolean expression $B_i$ within $S$ is called a
\emph{guard} and the construct $B_i \ra S_i$ is called a \emph{guarded
  command}.  

The symbol $\sep$ represents a nondeterministic choice between the
guarded commands $B_i \ra S_i$.
The alternative command
\[ \IFP \]
is executed by executing a statement $S_i$ for which the corresponding
guard $B_i$ evaluates to true. There is no rule saying which
statement among those whose guard evaluates to true should be selected.
If all guards $B_i$ evaluate to false, the alternative command \emph{aborts}. 

The selection of guarded commands in the context of a repetitive command
\[ \DOP \]
\NI
is performed in a similar way.
The difference is that after termination of a selected statement
$S_i$ the whole command is repeated starting with a new evaluation
of the guards $B_i$.
Moreover, in contrast to the alternative command, the repetitive command
properly terminates when all guards evaluate to false.

Dijkstra did not establish the notation of guarded commands
directly. Two earlier EWDs reveal that he first considered other
options, also about their intended semantics.

In
\href{https://www.cs.utexas.edu/users/EWD/ewd03xx/EWD398.PDF}{EWD398}
\cite{Dij73} he first used `,' to separate guarded commands, but
changed it halfway to $\sep$, reporting criticism of Don Knuth and
stating ``this whole report \underline{is} an experiment in
notation!''  Also, he wrote $B: S$ instead of $B \to S$ adopted in
\cite{Dij75} and used by him thereafter.  Further, for the repetitive
command he wrote that

\begin{quote}
  In the case of more than one executable command, it is again
  undefined which one will be selected, we postulate, however, that
  then they will be selected in ``fair random order'', i.e. we disallow
  the non-determinacy permanent neglect of a permanently executable
  guarded command from the list.
\end{quote}

However, a day later, in
\href{https://www.cs.utexas.edu/users/EWD/ewd03xx/EWD399.PDF}{EWD399}
\cite{Dij73a}, he admitted that this decision

\begin{quote}
  [$\dots$] was a mistake: for such constructs we prefer now not to
  exclude non-termination.  It is just too tricky if the termination
  —and in particular: the proof of the termination--- has to rely on
  the fair randomness of the selection and we had better restrict
  ourselves to constructs w[h]ere each guarded command, when executed,
  implies a further approaching of the terminal state.
\end{quote}

We shall return to this problem of fairness shortly. But first let us
focus on the main feature of guarded commands, the nondeterminism they
introduce.  However, this nondeterminism is of a different type than
the one we discussed so far: by definition a guarded command program
establishes the desired postcondition if \emph{all} possible
executions establish it. This kind of nondeterminism was later termed
\emph{demonic nondeterminism}.

This seems at the first sight like a flawed decision: why
should one complicate the matters by adding to the program more
possible executions paths, when one will suffice? But, as we shall
soon see, there are good reasons for doing it.

An often cited example in favour of the guarded commands language
is the formalization of Euclid's algorithm that computes the
greatest common divisor (\emph{gcd}) of two positive integers $x$ and $y$
\[\DO x>y \ra x:=x-y \ \sep \ x<y \ra y:=y-x\ \OD \]
that terminates with the \emph{gcd} of the initial values of $x$ and $y$
equal to their final, common, value.

However, this program is not nondeterministic: for any initial value
of the variables $x$ and $y$ there is only one possible program
execution. This program actually illustrates something else: that
guarded commands allow one to write more elegant algorithms.  Here the
variables $x$ and $y$ are treated symmetrically which is not the case
when a deterministic program is used.

In another simple example from Dijkstra's paper one is asked to
compute the maximum $max$ of two numbers, $x$ and $y$:
\[\IF x \geq y \ra max:=x \ \sep \ y \geq x \ra max:=y \ \IF \]
It illustrates the nondeterminism in a mildest possible form: when
$x=y$ two executions are possible, but the outcome is still the
same.

A slightly more involved example of a nondeterministic program with a
deterministic outcome is Dijkstra's solution to the following
problem: assign to the variables $x_1, x_2, x_3$, and $x_4$ an
ordered permutation of the values $X_1, X_2, X_3$, and $X_4$, i.e.,
one such that $x_1 \leq x_2 \leq x_3 \leq x_4$ holds.  The program
uses a parallel assignment that forms part of the guarded commands
language:

\begin{mytabbing}
\qquad\qquad
\= $x_1, x_2, x_3, x_4 := X_1, X_2, X_3, X_4$; \\
\> \DO       \= $x_1 > x_2 \ra x_1, x_2 := x_2, x_1$ \\
\> $\sep$    \> $x_2 > x_3 \ra x_2, x_3 := x_3, x_2$ \\
\> $\sep$    \>  $x_3 > x_4 \ra x_3, x_4 := x_4, x_3$ \\
\> \OD
\end{mytabbing}

Upon exit all guards evaluate to false, i.e.,
$x_1 \leq x_2 \leq x_3 \leq x_4$ holds, as desired.  The relevant
invariant is that $x_1, x_2, x_3, x_4$ is a permutation of $X_1, X_2, X_3, X_4$,

Finally, the following example of Dijkstra results in a program with a
nondeterministic outcome. The problem is to find for a fixed $n > 0$
and a fixed integer-valued function $f$ defined on $\{0, \LL, n-1\}$ 
a maximum point of $f$, i.e., a value $k$ such that
\[
  k \in \{0, \LL, n-1\} \land \fa i \in \{0, \LL, n-1\}: f(k) \geq f(i).
\]
A simple solution is the following program:

\begin{mytabbing}
\qquad\qquad
\= $k :=0;  j:= 1$; \\
\> \DO  $j \neq n$ \ra \= \IF    $f(j) \leq f(k) \ra j:=j+1$ \\
\>                     \> $\sep$ $f(j) \geq f(k) \ra k:=j; \ j:=j+1$ \\
\>                     \> \FI \\
\> \OD
\end{mytabbing}

\noindent
It scans the values of $f$ starting with the argument 0, updates the
value of $k$ in case a new maximum is found (when $f(j) < f(k)$),
and optionally updates the value of $k$ in case another current
maximum is found (when $f(j) = f(k)$).
The relevant invariant is here
\[
  k \in \{0, \LL, j-1\} \land j \leq n \land \fa i \in \{0, \LL, j-1\}: f(k) \geq f(i).
\]
Indeed, it is established by the initial assignment, maintained by each
loop iteration, and upon termination it implies the desired condition on $k$.
Note that the program can compute in $k$ 
any maximum point of $f$.

All these small examples (and there are no others in Dijkstra's paper)
do not provide convincing reasons for embracing nondeterminism
provided by the guarded commands language. A year after \cite{Dij75}
appeared, Dijkstra published his book \cite{Dij76} in which he derived
several elegant guarded command programs, including a more efficient version
  of the above program which avoids the recomputation of $f(j)$ and $f(k)$.
But inspecting these programs we found only a few examples in which
guards were not mutually exclusive and only two programs with a
nondeterministic outcome.

So why then has demonic nondeterminism, as present in the guarded command
language, turned out to be so influential?  In what follows we discuss 
subsequent developments that provide some answers to this
question. Many accounts of the guarded command language discuss it, as
Dijkstra originally did, together with its weakest precondition
semantics.  But to appreciate the nondeterminism Dijkstra introduced
in our view it is useful to separate the language from its weakest
precondition semantics.

\section{Some considerations on guarded commands}

Dijkstra's famous article \emph{Go To Statement Considered Harmful}
\cite{Dij68a} shows that he was aware of an ancestor of his
alternative command in the form of a conditional expression
$(B_1 \to e_1, \LL, B_n \to e_n)$ introduced by John McCarthy in
\cite{McC63}. Here $B_is$ are Boolean expressions and $e_is$ are
expressions.  The Boolean expressions do not need to be mutually
exclusive, but the conditional expressions are deterministic: when
executed the $(B_1 \to e_1, \LL B_n \to e_n)$ yields the value of the
first expression $e_i$ for which $B_i$ evaluates to true.  When all
$B_is$ evaluate to false, the conditional expression is supposed to be
undefined. So $(B_1 \to e_1, \LL B_n \to e_n)$ is a shorthand for a
nested \textbf{if}-\textbf{then}-\textbf{else} statement.

Dijkstra's explicit introduction of an abort, as opposed to McCarthy's
reference to `undefined', is useful because it provides a simple way of
implementing an \textbf{assert} $B$ statement that checks whether
assertion $B$ holds and causes an abort when this is not the case.

McCarthy worked within the framework of a functional language, so he
was constrained to use recursion instead of a looping construct.  As a
result Euclid's algorithm is formalized in his notation as follows:
\[
  gcd(m,n) = (m > n \to gcd(m-n,n), \ n > m \to gcd(m,n-m), \ m=n \to m),
\]
which is less elegant than Dijkstra's solution, due to the need for the final
component of the conditional expression.

In contrast, as already explained, Dijkstra did not prescribe any
order in which the guards are selected and ensured that his
repetitive command was not defined using the alternative
command.  In \cite{Dij75} he motivated his introduction of
nondeterminism (called by him nondeterminacy) as follows:

\begin{quote}
Having worked mainly with hardly self-checking
hardware, with which nonreproducing behavior of user programs is a
very strong indication of a machine malfunctioning, I had to overcome
a considerable mental resistance before I found myself willing to
consider nondeterministic programs seriously. [$\dots$]
Whether nondeterminacy is eventually removed mechanically---in
order not to mislead the maintenance engineer---or (perhaps only
partly) by the programmer himself because, at second thought, he does
care---e.g, for reasons of efficiency---which alternative is chosen is
something I leave entirely to the circumstances. In any case we can
appreciate the nondeterministic program as a helpful stepping stone.  
\end{quote}

But soon he overcame this resistance and one year later he wrote:
\begin{quote}
  Eventually, I came to regard nondeterminacy as the normal situation,
  determinacy being reduced to a ---not even very interesting--- special
  case. \cite[page xv]{Dij76}
\end{quote}

McCarthy's semantics of conditional expression can be viewed as an
example of such a `mechanical removal' of nondeterminism. However,
keeping nondeterminism intact often leads to simpler and more natural
programs even if the outcome is deterministic.  In some programs the
considered alternatives do not need to be mutually exclusive as long
as all cases are covered.

A beautiful example was provided by David Gries in his book
\cite{Gri81}.  Consider the following problem due to Wim Feijen. (We
follow here the presentation of Gries.)

Given are three magnetic tapes, each containing a list of different
names in alphabetical order.  The first contains the names of people
working at IBM Yorktown Heights, the second the names of students at
Columbia University and the third the names of people on welfare in
New York City.  It is known that at least one person is on all three
lists.  The problem is to locate the alphabetically first such person.

In \cite{Gri81} the following elegant program solving this problem was
systematically derived.  We assume here that the lists of names are
given in the form of ordered arrays $a[0:M], b[0:M]$, and $c[0:M]$:

\begin{mytabbing}
\qquad\qquad 
\= $i:=0;\ j:=0;\ k:=0;$                                 \\
\> \DO      \= $a[i]<b[j] \ra$ \= $i:=i+1$                              \\
\> $\sep$   \> $b[j]<c[k] \ra$ \> $j:=j+1$                              \\
\> $\sep$   \> $c[k]<a[i] \ra$ \> $k:=k+1$                              \\
\> \OD                                                         
\end{mytabbing}

Note that upon termination of the loop $a[i] = b[j] = c[k]$ holds.
The appropriate invariant is
\[
  0 \leq i \leq i_0 \A 0 \leq j \leq j_0 \A 0 \leq k \leq k_0 \A r
\]
where $r$ states that the arrays
$a[0:M], b[0:M]$, and $c[0:M]$ are ordered, $i_0, j_0, k_0 \leq M$
and 
$(i_0,j_0,k_0)$ is the lexicographically smallest triple such that
$a[i_0] = b[j_0] = c[k_0]$.

This program uses nondeterministic guards, so various computations are
possible. Still, it has a deterministic outcome.

In general, as soon as two or more guards are used in a loop, in the
customary, deterministic, version of the program one is forced to use
a, possibly nested, \textbf{if}-\textbf{then}-\textbf{else} statement,
like in McCarthy's `determinisation' approach. It imposes an
evaluation order of the guards, destroys symmetry between them, and
does not make the resulting programs easier to verify.

\section{Modeling parallel programs}
\label{sec:parallel}

Concurrent programs, introduced in Dijkstra's \emph{Cooperating sequential
processes} paper \cite{Dij68b}, can share variables, which makes it
difficult to reason about them. Therefore, starting with \cite{AM71}
and \cite{FS78,FS81}, various authors proposed to analyze them at the
level of nondeterministic programs, where the nondeterminism reflects
existence of various component programs. Such a reduction is possible
if one assumes that no concurrent reading and writing of variables
takes place.

Using guarded commands it is possible to make the link between these
two classes of programs explicit by a transformation.  The precise
transformation is a bit laborious, see \cite{FS78}, so we illustrate
it by an example taken from \cite{ABO09}. Consider the following
concurrent program 
due to Owicki and Gries [\citeyear{OG76a}] that searches for a positive
value in an integer array $ia[0:N]$:

\begin{mytabbing}
\qquad\qquad \= $i:=1;\ j:=2;\ oddtop:=N+1;\ eventop:=N+1$;       \\
            \> $[S_1 \| S_2]$;                                    \\
\> $k:=min(oddtop,eventop)$,
\end{mytabbing}

\NI
where $S_1$ and $S_2$ are
deterministic components $S_1$ and $S_2$ scanning the odd and the even
subscripts of $ia$, respectively:

\begin{mytabbing}
\qquad\qquad $S_1 \equiv$ 
$a$: \= \WD{i<min(oddtop,eventop)}                            \\
\> \qquad $b$: \IF $ia[i]>0$ \= \THEN \= $c$: $oddtop:=i$ 
                             \ELSE \= $d$: $i:=i+2$\ \FI      \\
\> \OD                                                         
\end{mytabbing}
and
\begin{mytabbing}
\qquad\qquad $S_2 \equiv$ 
$a$: \= \WD{j<min(oddtop,eventop)}                               \\                                                    
\> \qquad $b$: \IF $ia[j]>0$ \= \THEN \= $c$: $eventop:=j$        
                             \ELSE \= $d$: $j:=j+2$\ \FI         \\
\> \OD                                                          
\end{mytabbing}

Upon termination of both components, the minimum of two shared integer variables
$oddtop$ and $eventop$ is checked.
The labels $a, b, c, d$, and $e$
are added here to clarify the transformation.
The parallel composition  $S \equiv [S_1 \|\, S_2]$ is
transformed into the following guarded commands program $T(S)$ 
with a single repetitive command that employs the
control variables $cv_1$ and $cv_2$ for $S_1$ and $S_2$
that can assume the values of the labels: 
\begin{mytabbing}
\qquad\qquad $T(S)\ \equiv\ $
\= $cv_1:=a;\ cv_2:=a;$                                                 \\
\> \DO \= $cv_1=a \A i<min(oddtop,eventop) \ra cv_1:=b$      \+         \\
        $\sep$ \> $cv_1=a \A \neg (i<min(oddtop,eventop)) \ra cv_1:=e$  \\
        $\sep$ \> $cv_1=b \A ia[i]>0 \ra cv_1:=c$                        \\
        $\sep$ \> $cv_1=b \A \neg (ia[i]>0) \ra cv_1:=d$                 \\
        $\sep$ \> $cv_1=c \ra oddtop:=i;\ cv_1:=a$                      \\
        $\sep$ \> $cv_1=d \ra i:=i+2;\ cv_1:=a$                         \\
        $\sep$ \> $cv_2=a \A j<min(oddtop,eventop) \ra cv_2:=b$         \\
        $\sep$ \> $cv_2=a \A \neg (j<min(oddtop,eventop)) \ra cv_2:=e$  \\
        $\sep$ \> $cv_2=b \A ia[j]>0 \ra cv_2:=c$                        \\
        $\sep$ \> $cv_2=b \A \neg (ia[j]>0) \ra cv_2:=d$                 \\
        $\sep$ \> $cv_2=c \ra eventop:=j;\ cv_2:=a$                     \\
        $\sep$ \> $cv_2=d \ra j:=j+2;\ cv_2:=a$               \-        \\
\> \OD; \\
\> \IF $cv_1=e \A cv_2=e \ra skip$ \FI 
\end{mytabbing}
Note that the repetitive command exhibits nondeterminism. For example,
when $cv_1 = cv_2 = a$, two guarded commands can be chosen next.  This
corresponds to the \emph{interleaving semantics} of concurrency that
we assume here. When the repetitive command has terminated, the final
alternative command checks whether this termination is the one
intended by the original concurrent program $S$. This is the case when
both $cv_1$ and $cv_2$ store the value $e$. In the current example,
this check is trivially satisfied and thus the alternative command
could be omitted.

However, for concurrent programs with synchronization primitives, a
termination of the repetitive command may be due to a deadlock in the
concurrent program.  Then the final alternative command is used to
transform the deadlock into a failure, indicating an undesirable state
at the level of nondeterministic programs.  For details of this
transformation we refer to Chapter 10 of \cite{ABO09}.

This transformation allows us to clarify that the nondeterminism
resulting from parallelism is the one used in the guarded commands
language.  However, this example also reveals a drawback of the
transformation: the structure of the original parallel program gets
lost.  The resulting nondeterministic program represents a single loop
at the level of an assembly language with atomic actions explicitly
listed.  The assignments to the control variables correspond to
\textbf{go to} statements, which explains why reasoning about the
resulting program is difficult.  Interestingly, this problem does not
arise for the transformation of the CSP programs that we give in the
next section.

\section{Communicating Sequential Processes and their relation to guarded commands}
\label{sec:CSP}

Dijkstra's quoted statement, ``In any case we can appreciate the
nondeterministic program as a helpful stepping stone'' suggests that
he envisaged some extensions of the guarded command language. But in
his book \cite{Dij76} he only augmented it with local variables by
providing an extensive notation for various uses of local and global
variables, and added arrays. In his subsequent research he only used
the resulting language.

However, his discussion of the guarded commands program formalizing
Euclid's algorithm suggests that he also envisaged some connection with
concurrency. He suggested that the program could be viewed as a
synchronization of two cyclic processes $\DO x:=x-y \ \OD$ and
$\DO y:=y-x \ \OD$ in such a way that the relation $x> 0 \land y>0$ is
kept invariantly true. Still, he did not pursue this idea
further.

Subsequent research showed that guarded command programs can be viewed
as a natural layer lying between deterministic and concurrent
programs. This was first made clear in 1978 by Tony Hoare who
introduced in \cite{Hoa78} an elegant language proposal for
distributed programming that he called Communicating Sequential
Processes (abbreviated to CSP) in clear reference to Dijkstra's
\emph{Cooperating sequential processes} paper \cite{Dij68b}.

Hoare stated seven essential aspects of his proposal, mentioning as
the first one Dijkstra's guarded commands ``as the sole means of
introducing and controlling nondeterminism''. The second one also
referred to Dijkstra, namely to his parallel command, according to
which, ``All the processes start simultaneously, and the parallel
command ends only when they are all finished.''
It is useful to discuss CSP in some detail to see how each of
these two aspects results in the same type of nondeterminism.

In Dijkstra's cooperating sequential processes model processes
communicate with each other by updating global variables. 
By contrast, in CSP processes communicate solely by means of the
\emph{input} and \emph{output} commands, which are atomic statements
that are executed in a synchronized fashion. So CSP processes do not
share variables.

For the purpose of communication CSP processes have names.  The input
command has the form $P?x$, which is a request to process (named) $P$
to provide a value to the variable $x$, while the output command has
the form $Q!t$, which is a granting of the value of the expression $t$
to process (named) $Q$. When the types of $x$ and $t$ match and the
processes refer to each other, we say that the considered input
and output commands \emph{correspond}. They are then executed
simultaneously; the effect is that of executing the assignment
$x:=t$.\footnote{Note that not all assignments can be modelled this
  way.  For instance, the assignment $x:=x+1$ cannot be reproduced
  since processes do not share variables.}

In CSP a single input command is also allowed to be part of a guard. The
restriction to input commands was dictated by implementation
considerations. But once it was clarified how to implement the use of
output commands in guards, they were admitted as part of a guard, as
well. So, in the sequel we admit both input and output commands in guards. 
Thus guards are of the form $B; \alpha$, where $B$ is a Boolean expression
and $\alpha$ is an input or output command, \emph{i/o command} for short.
We assume that such an extended guard \emph{fails} when the
Boolean part evaluates to false.\footnote{Hoare also allowed an extended
guard to fail when its i/o command refers to a process that terminated.
For simplicity do not adopt this assumption here.}

To illustrate the language consider an example, taken from
\cite{ABO09}, which is a modified version of an example given in
\cite{Hoa78}.  In what follows we refer to the repetitive commands of
CSP as \textbf{do} loops.

We wish to transmit a sequence of characters from the process \SENDER\
to the process \RECEIVER\ with all blank characters (represented by \BLANK)
deleted. To this end we employ an intermediary process \FILTER\ and
consider a distributed program
\[[\SENDER\ \|\ \FILTER\ \|\ \RECEIVER] \]

The sequence of characters is initially stored by the process \SENDER\
in the array $a[0:M-1]$ of characters, with \AST\ as the last
character. The process \FILTER\ uses an array $b[0:M-1]$ of characters
as an intermediate store for processing the character sequence and the
process \RECEIVER\ has an array $c[0:M-1]$ of the same type to store
the result of the filtering process.  For coordinating its activities
the process \FILTER\ uses two integer variables $in$ and $out$
pointing to elements in the array $a[0:M-1]$.
The processes are defined as follows:

\begin{mytabbing}
\qquad
\= $\RECEIVER ::$ \=                                                \kill
\> $\SENDER ::$  \> $i:=0$;  $\DO i \neq M;\FILTER\ !\,a[i] \ra i:=i+1\ \OD$   \\
[\medskipamount]
\> $\FILTER\ ::$  \> $in:=0;\ out:=0;\ x:=\BLANK$;                   \\
\> \> \DO    \= $x \neq \AST; \SENDER\ ?\,x \ra$                               \\
\> \>        \> \qquad \= \IF    \= $x=\BLANK \ra$ \= $skip$            \\
\> \>        \>        \> $\sep$ \> $x \neq \BLANK \ra$ \> $b[in]:=x;\ in:=in+1$   \\
\> \>        \>        \> \FI                                           \\
\> \> $\sep$ \> $out \neq in; \RECEIVER\ !\,b[out] \ra out:=out+1$      \\
\> \> \OD     \\                                                        
[\medskipamount]
\> $\RECEIVER ::$ \> $j:=0;\ y:=\BLANK$;                            \\
\>                   \> $\DO y \neq \AST; \FILTER\ ?\,y \ra c[j]:=y;\ j:=j+1\ \OD$
\end{mytabbing}

Note that the processes \SENDER\ and \RECEIVER\ are deterministic, in
the sense that each extended guarded command either has just one guard
or the Boolean parts of the used guards are mutually exclusive
(this second case does not occur here), while
\FILTER\ is nondeterministic as it uses a \textbf{do} loop with two
extended guards the Boolean parts of which are not mutually
exclusive. They represent two possible actions for \FILTER: to
communicate with \SENDER\ or with \RECEIVER.

Hoare presented in his article several elegant examples of CSP
programs.  In some of them the processes are deterministic. But even
then, if there are four or more processes, the resulting program is
nondeterministic, since it admits more than one computation. 

By assumption a CSP program is correct if all of its computations
properly terminate in a state that satisfies the assumed
postcondition. So this is exactly demonic
nondeterminism, like in the case of parallel programs.

This makes it possible to translate CSP programs in a simple way to
guarded command programs.  In \cite{ABO09} we provided such a
transformation for a fragment of CSP, in which the above example is
written. The CSP programs in this fragment are of the form
\[ S \equiv \PP, \]
where each process $S_i$ is of the form
\[
  S_i \equiv S_{i,0};\ \DO \sep^{m_i}_{j=1}\ B_{i,j};\al_{i,j} \ra  S_{i,j}\ \OD, 
\]
each $S_{i,j}$ is a guarded command program, each $B_{i,j}$ is a
Boolean expression, and each $\al_{i,j}$ is an i/o command.  So each
process $S_i$ has a single \DO loop in which i/o commands appear only
in the guards. No further i/o commands are allowed in the
initialization part $S_{i,0}$ or in the bodies $S_{i,j}$ of the
guarded commands.

As shown in \cite{ABC87} and \cite{Zob88} each CSP
program can be transformed into a program in this fragment by
introducing some control variables.

As abbreviation we introduce
\[ \Gamma=\{(i,j,r,s) \mid \al_{i,j}\ \mbox{and}
  \ \al_{r,s}\ \mbox{correspond and}\ i<r\}.
\]

According to the CSP semantics two generalized guards from
different processes can be passed jointly when their i/o commands
correspond and their Boolean parts evaluate to true. Then the
communication between the i/o commands takes place.  The effect of a
communication between two corresponding i/o commands
$\al_1 \equiv P?x$ and $\al_2 \equiv Q!t$ is the assignment $x:=t$.
Formally, for two such commands we define
\[ \Eff(\al_1,\al_2) \equiv \Eff(\al_2,\al_1) \equiv x:=t. \]

We transform $S$ into the following guarded commands program $T(S)$:

\begin{mytabbing}
\qquad $T(S)\ \equiv\ $
\= $S_{1,0};\ \LL;\ S_{n,0};$                                   \\
\> $\DO \sep_{(i,j,r,s)\in\Gamma}\ B_{i,j} \A B_{r,s} \ra$
                \= $\Eff(\al_{i,j},\al_{r,s});$   \\
\>              \> $S_{i,j};\ S_{r,s}$ \\
\> $\OD$,
\end{mytabbing}
\NI
where we use of elements of $\Gamma$ to list all guards
in the \textbf{do} loop. In the degenerate case when $\Gamma=\ES$
we drop this loop from $T(S)$.

For example, for $SFR \equiv [\SENDER\, \|\, \FILTER\, \|\, \RECEIVER ]$ we obtain the following guarded commands program:

\begin{mytabbing}
\qquad $T(SFR)\ \equiv\ $
\= $i:=0;\ in:=0;\ out:=0;\ x:=\BLANK;\ j:=0;\ y:=\BLANK$; \\
 \> \DO    \= $i \neq M \land x \neq \AST \ra$\= $x:= a[i];\ i:= i+1;$ \\
 \>        \>                                 \> \IF \= $x=\BLANK \ra$ \= $skip$            \\
 \>        \>                                          \> $\sep$         \> $x \neq \BLANK \ra$ \> $b[in]:=x;\ in:=in+1$   \\
\>        \>                                  \> \FI                                           \\
\> $\sep$ \> $out \neq in \land y \neq \AST \ra y:= b[out];\ out:=out+1;\ c[j]:=y;\ j:=j+1$      \\
\> \OD                 
\end{mytabbing}

The semantics of $S$ and $T(S)$ are not identical because their termination
behavior is different. However, the final
states of properly terminating computations of $S$ and the final
states of properly terminating computations of $T(S)$ that satisfy
the condition
\[ \TERM
\equiv \bigwedge^n_{i=1}\ \bigwedge^{m_i}_{j=1}\ \neg B_{i,j} 
\]
coincide.  An interested reader can consult Chapter 11 of
\cite{ABO09}.

The above transformation makes precise what we already mentioned when
discussing an example CSP program: the CSP language introduces
nondeterminism in two ways. The first one comes from allowing guarded
commands; in the transformed program these are the programs $S_{a,b}$. The
second one results from synchronous communication, modelled in the
transformation by means of the outer repetitive command.  So both ways
are instances of demonic nondeterminism.  

Thanks to the special form of CSP programs the transformed
program does not introduce any new variables.  As a result this
transformation suggests a simple way to reason about
correctness of CSP programs, by considering the translated guarded
commands program, see Chapter 11 of \cite{ABO09}.

\section{Fairness}

One of the programs in \cite{Dij76} with a nondeterministic
outcome is the following one:

\begin{mytabbing}
\qquad\qquad 
\= $goon : = \mathbf{true}; \ x:=1;$                                 \\
\> \DO      \= $goon \ra$ \= $x:=x+1$                              \\
\> $\sep$   \> $goon \ra$ \= $goon:= \mathbf{false}$                              \\
\> \OD                                                         
\end{mytabbing}

The problem Dijkstra was addressing was that of writing a program that
sets $x$ to any natural number.  He concluded using his weakest
precondition semantics that no guarded commands program exists that
sets $x$ to any natural number \emph{and} always terminates.  Note
that the above program can set $x$ to an arbitrary natural number but
also can diverge.

As noted in \cite{Plo76},  Dijkstra's conclusion can be obtained in a
more direct way by appealing to any operational semantics that
formalizes the notion of a computation, by representing the
computations in the form of a tree. Branching models the execution of
a guarded command; each branch corresponds to a successful evaluation
of a guard. Given an input state of a guarded command program we have
then a finitely branching tree representing all possible computations.
Let us call it a \emph{computation tree}.

Denote a program that sets $x$ to any natural number by $x :=?$.
Suppose that it can be represented by a guarded commands program
$S$. Then for any input state the computations of $S$ form a finitely
branching computation tree with infinitely many leaves, each of them
corresponding to a different natural number assigned to $x$.  We can
now appeal to K\"{o}nig's Lemma which states that any finitely
branching infinite tree has an infinite path \cite{Kon27}. It
implies that in every computation tree of $S$ an infinite computation
exists, i.e., that for every input state the program $S$ can diverge.

Dijkstra's proof contained in Chapter 9 of \cite{Dij76} proceeds
differently.  He first showed that his predicate transformer \emph{wp}
is continuous in the predicate argument and then that the
specification of $x :=?$ in terms of \emph{wp} contradicts continuity.
In his terminology, the program $x :=?$ introduces \emph{unbounded
  nondeterminism}, which means that no a priori upper bound for the
final value of $x$ can be given.

The program $x :=?$ occupied his attention a number of times.  In
EWD673 published as \cite{EWD:EWD673pub} he noted that to prove
termination of guarded command programs augmented with the program
$x :=?$ it does not suffice to use integer-valued bound functions and
one has to resort to well-founded relations.  In turn, in EWD675
published as \cite{EWD:EWD675pub} he noticed that the converse
implication holds, as well: the existence of a program for which
\emph{wp} is not continuous implies the existence of a program with
unbounded nondeterminism.

Soon more in-depth studies of the program $x:=?$ and the consequences
of its addition to deterministic programs followed. 
In particular, Chandra showed in \cite{Cha78} that the halting problem
for programs admitting $x:=?$ is $\Pi^1_1$ complete, so
of higher complexity than the customary halting problem for computable functions 
(see also \cite{STB89} for a further discussion of this problem), while 
Back [\citeyear{Bac80}] and Boom [\citeyear{Boo82}] advocated use of $x:=?$ as a
convenient form of program abstraction that deliberately ignores
the details of an implementation.
One may note here that Hilbert's $\epsilon$ notation essentially
serves a similar purpose: $\epsilon\, x\, \phi$ picks an $x$ that
satisfies the formula $\phi$.
We wish to remain within the
realm of guarded commands, so limit ourselves to a clarification of
the relation of the program $x:=?$ with the notion of fairness.

If we assume that the guards are selected in `fair random order',
then the program given at the beginning of this section always terminates
and can set $x$ to any natural number.
Here fairness refers to \emph{weak fairness}, or \emph{justice}, which requires that
each guard that continuously evaluates to true is eventually chosen.
 So the assumption of weak fairness for guarded commands allows
us to implement the program $x :=?$.

As shown in \cite{Boo82} and independently, though later, in \cite{AO83}
the converse holds, as well. We follow here a presentation from the latter paper.
We call a guarded commands program \emph{deterministic} if in each
alternative or repetitive command used in it the guards are mutually exclusive.
We call a guarded commands program
\[ 
S \equiv S_0;\ \DOPa, 
\]
\emph{one-level nondeterministic} if $S_0,S_1,\LL,S_n$ are
deterministic programs.

Given now such a one-level nondeterministic program
we transform it into the following guarded commands program that uses the $x:=?$ programs:

\begin{mytabbing}
\qquad\qquad
$T_{wf}(S)\ \equiv\ $
\= $S_0;\ z_1:=?;\ \LL;\ z_n:=?;$                               \\
\> $\DO \sep^n_{i=1}\ B_i \A z_i=min \ \C{z_k \mid 1 \leq k \leq n} \to$        \\
\> \qquad\qquad \= $z_i:=?$;                                    \\
\>              \> {\bf for all} $j \in \C{\LLn} \setminus \C{i} \ \DO$    \\
\>              \> \qquad \ITE{B_j}{z_j:=z_j-1}{z_j:=?}                  \\
\>              \> \OD;                                         \\
\>              \> $S_i$                                        \\
\> \OD,
\end{mytabbing}
where $z_1,\LL,z_n$ are integer variables that do not occur in $S$.

Intuitively, the variables $z_1,\LL,z_n$ represent priorities
assigned to the $n$ guarded commands in the repetitive command of $S$.
A guarded command $i$ has higher priority than a command $j$ if $z_i<z_j$.
Call a guarded command \emph{enabled} if its guard evaluates to true.

Initially, the commands are assigned arbitrary priorities.  During
each iteration of the transformed repetitive command an enabled
command with the maximal priority, i.e., with the minimum value of
$z_i$, is selected.  Subsequently, its priority gets reset
arbitrarily, while the priorities of other commands are appropriately
modified: if the command is enabled then its priority gets increased
and otherwise it gets reset arbitrarily.  The idea is that by
repeatedly increasing the priority a continuously enabled guarded
command we ensure that it will be eventually selected.

This way the transformation models weak fairness. More precisely, 
if we ignore the values of the variables $z_1,\LL,z_n$, the
computations of $T_{wf}(S)$ are exactly the weakly fair computations of
$S$. A similar transformation can be shown to model
a more demanding form of fairness (called \emph{strong fairness} or \emph{compassion})
according to which each guard that infinitely often evaluates to true
is also infinitely often selected.  An interested reader can consult
\cite{AO83} or Chapter 12 of \cite{ABO09}.

But are there some interesting guarded command programs for which the
assumption of fairness is of relevance? The answer is `yes'.  A nice
example was provided to us some time ago by Patrick Cousot who pointed
out that a crucial algorithm in their landmark paper \cite{CC77} which
introduced the idea of abstract interpretation relies on fairness. The
authors were interested in computing a least fixed point of a
monotonic operator in an asynchronous way by means of so-called
chaotic iterations.

Recall that a \emph{partial order} is a pair
$(A , \po )$ consisting of a set $A$ and a reflexive, antisymmetric
and transitive relation $\po$ on $A$.
Consider the $n$-fold Cartesian product $A^n$ of $A$ for some
$n \geq 2$ and extend the relation $\po$ componentwise from $A$ to
$A^n$. Then $(A^n, \po)$ is a partial order.
% for which
% \[
%   \mbox{$\bx \spo \by$ iff $\bx \po \by$ and $\bx \neq \by$},
% \]
% where $\bx \equiv (x_1,\LL,x_n)$ and $\by \equiv (y_1,\LL,y_n)$.

Next, consider a function
\[ 
  F:A^n \to A^n,
\]
and the $i$th component functions $F_i:A^n \to A$, where $i \in \{1, \LL, n\}$, each defined by
\[ 
  F_i(\bx)=y_i \ {\rm iff} \ F(\bx) = \by. 
\]
Suppose now that $F$ is \emph{monotonic}, that is, whenever
$ \bx \po \by$ then $F(\bx) \po F(\by)$.  Then the functions $F_i$ are
monotonic, as well.

Further, assume that $A^n$ is finite and has the $\po$-least element that we denote by $\ES$.
By the Knaster and Tarski  Theorem \cite{Tar55}
$F$ has a \emph{$\po$-least fixed point} $\mu F \in A^n$.
As in \cite{CC77} we wish to compute $\mu F$ \emph{asynchronously}. This is achieved by means
the following guarded commands program:

\begin{mytabbing}
\qquad\qquad 
\=  $\bx := \ES$; \\
\>  $\DO \sep^n_{i=1}\  \bx \neq F(\bx) \to x_i:=F_i(\bx) \ \OD$
\end{mytabbing}

This program can diverge, but it always terminates under the assumption
of weak fairness.  This is a consequence of a more general theorem
proved in \cite{CC77}. An assertional proof of
correctness of this program under the fairness assumption is given in
\cite{ABO09}.

Dijkstra had an ambiguous attitude to fairness. As noted  in Chapter \cite{Eme22}
% written by Allen Emerson,
of the book \cite{AH22}
Dijkstra stated in his
\href{https://www.cs.utexas.edu/users/EWD/ewd03xx/EWD310.PDF}{EWD310},
that appeared as \cite{Dij71}, that sequential
processes forming a parallel program should ``proceed with speed ratios,
unknown but for the fact that the speed ratios would differ from
zero'' and referred to this property as `fairness'.

In EWD391 dating from 1973 and published as \cite{EWD:EWD391pub},
when introducing self-stabilization he wrote:
\begin{quote}
In the middle of the ring stands a demon, each time giving, in ``fair
random order'' one of the machines the command ``to adjust itself''. (In
``fair random order'' means that in each infinite sequence of successive
commands issued by the d[a]emon, each machine has received the command to
adjust itself infinitely often.)
\end{quote}

In the two-page journal publication \cite{Dij74} that soon followed
the qualification `fair random order' disappeared, but `daemon' that
ensures it remained:
\begin{quote}
  In order to model the undefined speed ratios of the various
  machines, we introduce a central daemon [$\dots$].
\end{quote}

However, as we have seen when discussing the origin of guarded
commands, he rejected in EWD399 \cite{Dij73a} fairness at the level of
guarded commands.  In
\href{https://www.cs.utexas.edu/users/EWD/ewd07xx/EWD798.PDF}{EWD798}
\cite{EWD:EWD798} one can find the following revealing comment:
\begin{quote}
  David Park (Warwick University) spoke as a last minute replacement
  for Dana Scott on ``Fairness''. The talk was well-prepared and
  carefully delivered, but I don't care very much for the
  topic.
\end{quote}

Further, in his
\href{https://www.cs.utexas.edu/users/EWD/ewd10xx/EWD1013.PDF}{EWD1013}
\cite{Dij87}, titled \emph{Position paper on ``fairness''}, he plainly
turned against fairness and ended his informal discussion by
stating that ``My conclusion [$\dots$] is that fairness, being an
unworkable notion, can be ignored with impunity.''

It is easy to check that the transformation given in
Section~\ref{sec:parallel} translates fairness for parallel programs
assumed in \cite{Dij71} to weak fairness for guarded commands.  We are
bound to conclude that Dijkstra's opinions on fairness were not
consistent over the years.

At the time Dijkstra wrote his controversial note \cite{Dij87}
fairness was an accepted and a well-studied concept, see, e.g.,
\cite{Fra86}. The note did not change researchers' perception. It was
soon criticized, in particular by Leslie Lamport and Fred Schneider
who concluded in \cite{LS88} that Dijkstra's arguments against
fairness apply equally well against termination, or more generally,
against any liveness notion.

\section{Nondeterminism: further developments}

Dijkstra's guarded commands language was not the last word on
nondeterminism in computer science. Subsequent developments, to which
he did not contribute, brought new insights, notably by clarifying the
consequences of both angelic and demonic nondeterminism in the context
of parallelism. In what follows we provide a short account of this
subject.

\subsection*{Taxonomy of nondeterminism}

Among several papers dealing with nondeterminism in the wake of
Dijkstra's guarded commands we would like to single out two.  In
\cite{HP78} Harel and Pratt investigated nondeterministic programs in
the context of Dynamic Logic and related their work to the weakest
precondition approach of Dijkstra. In their approach the programs were
built up from assignments to simple variables using a set of basic
operators: $\cup$ for nondeterministic choice and $;$ for sequential
composition, $B?$ for testing a Boolean condition $B$, and $^*$ for
iteration.  An alternative command $\IFPa$ can be viewed as an
abbreviation for the program $\cup_{i=1}^n (B_i?;S_i)$ and a repetitive
command $\DOPa$ an abbreviation for the program
$((\bigvee_{i=1}^n B_i?); (\cup_{i=1}^n (B_i?;S_i)))^*; \neg (\bigvee_{i=1}^n B_i?)$,
see, e.g., \cite{Bak80}.

Semantically, each
program denotes a binary relation on states, augmented with the
symbols $\bot$ representing \emph{divergence} (nonterminating
computations) and $f$ representing \emph{failure}, i.e., a test
evaluating to false without having any immediate alternative to
pursue.  Such a relation describes the input-output behavior of a
given program.  It is defined by induction on the structure of
programs.  For example, the relation associated with the program
$S_1 \cup S_2$ is the union of the relations associated with the
programs $S_1$ and $S_2$.

The input/output relation does not describe how it is computed in a step-by-step manner. 
When executed in a given state, the program $S_1 \cup S_2$
chooses either $S_1$ or $S_2$ to compute the successor state.
For a given initial state, these nondeterministic choices can be 
systematically represented in a \emph{computation tree}.
Harel and Pratt distinguished four methods how such a computation tree
can be traversed:
(1) depth first,
(2) depth first with backtracking when a failure state in encountered,
(3) breadth first,
(4) breadth first combined with ignoring failure states.
For each method, they showed how to express the notion of total
correctness in dynamic logic. 

In the context of algebraic specifications of programming languages, 
Broy and Wirsing considered different kinds of nondeterminism  in their paper \cite{BW81}.
They called them:
(1) \emph{backtrack nondeterminism},
(2) \emph{choice nondeterminism},
(3) \emph{unbounded nondeterminism}, and
(4) \emph{loose nondeterminism}.
Option~(1) computes the whole set of possible outcomes, where any possibility
of nontermination must be taken. The choice of the output comes `after' the
computation of the set of all possible outputs.
Option~(2) corresponds to choices `during' the execution of alternative statements.
Option~(3) applies `prophetic' choices during the computation to avoid any
nonterminating computations, thereby typically creating unboundedly many
good outcomes. 
Finally, option~(4) corresponds to choices `before' the execution of the program.

Broy and Wirsing were also early users of the terminology of
\emph{angelic} nondeterminism, which they identified with~(3),
\emph{demonic} nondeterminism, which they  identified with~(1), and
\emph{erratic} nondeterminism, which they  identified with~(2).
We could not trace who first introduced this terminology, though it has been often attributed to Tony Hoare.

\subsection*{Nondeterminism in a context}

\begin{figure}[htbp]

\begin{center} 

\begin{tikzpicture}[transitionsystem]

\node[state] (p1) {$p_1$};
\node[above=of p1] (p0) {$P:$};
\node[state, right=of p1] (p2) {$p_2$};
\node[state, above right=of p2] (p3) {$p_3$};
\node[state, below right=of p2] (p4) {$p_4$};

\node[state, below right=of p3] (q1) {$q_1$};
\node[above=of q1] (q0) {$Q:$};
\node[state, above right=of q1] (q2) {$q_2$};
\node[state, below right=of q1] (q3) {$q_2'$};
\node[state, right=of q2] (q4) {$q_3$};
\node[state, right=of q3] (q5) {$q_4$};

\path
(p1) edge node {$i$} (p2)
(p2) edge node {$t$} (p3)
(p2) edge[swap] node {$c$} (p4)
;

\path
(q1) edge node {$i$} (q2)
(q1) edge[swap] node {$i$} (q3)
(q2) edge node {$t$} (q4)
(q3) edge[swap] node {$c$} (q5)
;

\end{tikzpicture}

\end{center}
\caption{\label{fig:Milner} 
Two automata, $P$ being deterministic
and $Q$ being nondeterministic on input of $i$, accept the same
language consisting of the words $i \, t$ and $i \, c$.
However, when viewed as processes interacting with a user, they are different.
Suppose the process models a vending machine, $i$ corresponds to the user's 
action of inserting a coin into the machine,
and $t$ and $c$ to the user's choice of tea or coffee. 
Then, after insertion of the coin, $P$ is in state $p_2$ and
offers both tea and coffee to the user.
However, $Q$ makes a tacit choice by moving either to state $q_2$
or to state $q_2'$ after a coin is inserted. In state $q_2$ it offers only
tea, and in state $q_2'$ only coffee, never both.
Thus, from the user's perspective, the deterministic automaton is better
because when using it, no decision is taken without consulting her or him.}
\end{figure}
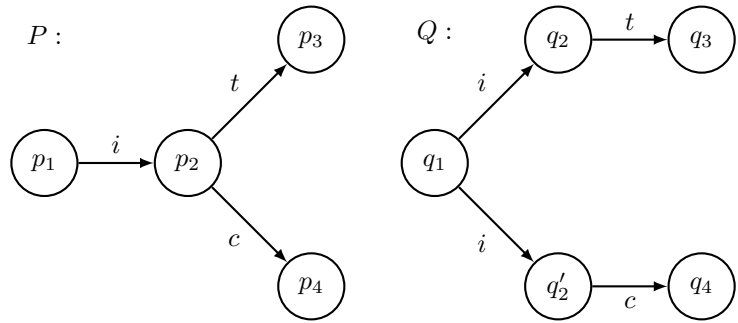

In his books on CCS and the $\Pi$-calculus \cite{Mil80,Mil99},
Robin Milner gave a simple example of two finite automata,
one deterministic and one nondeterministic, that are equivalent when
the accepted languages are compared. However, Milner argued that
they are essentially different, when they are considered as processes
interacting with a user or an environment. The essence of the example is shown
in Figure~\ref{fig:Milner}, adapted from \cite{Mil99}. Milner took this observation
as a motivation to develop a new notion of equivalence between processes, 
called \emph{bisimilarity} and based on the following notion of bisimulation, 
which is sensitive to nondeterminism.

Processes are like nondeterministic automata, 
with states and transitions between states that are labeled
by action symbols. We write $p \stackrel{a}{\longrightarrow} q$
for a transition from a state $p$ to a state $q$ labeled by $a$.
A process has an initial state and may have infinitely many states and thus transitions.

A \emph{bisimulation} between processes $P$ and $Q$ is a binary relation $\mathcal{R}$
between the states of $P$ and $Q$ such that 
whenever $p \mathcal{R} q$ holds, 
then every transition from $p$ can be simulated by a transition from $q$ with the
same label, such that the successor states are again in the relation $\mathcal{R}$,
and vice versa, every transition from $q$ can be simulated by a transition from $p$ with the
same label such that the successor states are again in the relation $\mathcal{R}$.
Processes are called \emph{bisimilar} if there exists a bisimulation
relating the initial states of the processes.

The processes shown in Figure~\ref{fig:Milner} are \emph{not} bisimilar.
Indeed, suppose that $\mathcal{R}$ is a bisimulation with $p_1 \mathcal{R} q_1$. Then the 
transition $p_1 \stackrel{i}{\longrightarrow} p_2$ can be simulated only by
$q_1 \stackrel{i}{\longrightarrow} q_2$, which implies $p_2 \mathcal{R} q_2$.
However, now the transition $p_2 \stackrel{c}{\longrightarrow} p_4$ cannot
be simulated  from $q_2$ because there is no transition with label $c$.
Contradiction.
This formalizes the intuition that only process $P$ offers both tea and coffee,
whereas $Q$ offers either tea or coffee.

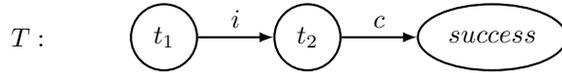
\begin{figure}[htbp]

\begin{center} 

\begin{tikzpicture}[transitionsystem]

\node[state] (t1) {$t_1$};
\node[left=of t1] (t0) {$T:$};
\node[state, right=of t1] (t2) {$t_2$};
\node[state, right=of t2] (t3) {$success$};

\path
(t1) edge node {$i$} (t2)
(t2) edge node {$c$} (t3)
;

\end{tikzpicture}

\end{center}
\caption{\label{fig:Test} 
This test $T$ checks whether a process can engage in first $i$ and then $c$.
}
\end{figure}

This new notion of equivalence triggered a copious research activity
resulting in various process equivalences that are sensitive to
nondeterminism but differing in various other aspects, see for example
\cite{Gla01}.  Of particular interest is the idea of \emph{testing}
processes due to De Nicola and Hennessy [\citeyear{NicolaH84,Hen88}].
In these works, the interaction of a (nondeterministic) process and a
user is explicitly formalized using a synchronous parallel
composition.  The user is formalized by a \emph{test}, which is a
process with some states marked as a \emph{success}. For an example
see Figure~\ref{fig:Test}.  The authors distinguish between two
options: a process may or must pass a test.  A process $P$ \emph{may}
pass a test $T$ if in \emph{some} maximal parallel computation with
$P$, synchronizing on transitions with the same label, the test $T$
reaches a \emph{success} state.  A process $P$ \emph{must} pass a test
$T$ if in \emph{all} such computations the test $T$ reaches a
\emph{success} state.

This leads to \emph{may} and \emph{must} equivalences.
Two processes are \emph{may} equivalent if each test that one process
may pass the other may pass as well, and analogously for the
\emph{must} equivalence.
So \emph{may} equivalence corresponds to angelic nondeterminism,
and \emph{must} equivalence to demonic nondeterminism.

As an example, consider the processes $P$ and $Q$ from
Figure~\ref{fig:Milner} and the test $T$ from Figure~\ref{fig:Test}.
Then $P$ both may and must pass the test $T$, whereas $Q$ only may
pass it because in a synchronous parallel computation it can get stuck
in the state pair $(q_2, t_2)$, without being able to reach
\emph{success}.  So in this simple example, both bisimilarity and
\emph{must} equivalence reveal the same difference between the
deterministic process $P$ and the nondeterministic process $Q$.  In
general, bisimilarity is finer than the testing equivalences, see
again \cite{Gla01}.

Also CSP, originally built on Dijkstra's guarded commands as explained
in Section~\ref{sec:CSP}, was developed further into a 
more algebraically oriented language that for clarity we call
here `new CSP'. It is described in Hoare's book \cite{Hoa85}.
While guarded commands and the original CSP were notationally close
to programming language constructs, where the nondeterminism appears
only within the alternative command or the \DO loop,
the new CSP introduced separate operators for each concept of the language. 
These can be freely combined to build up processes.
We focus here on two nondeterministic operators introduced by Hoare.

\emph{Internal nondeterminism} is denoted by the binary operator $\sqcap$,
called \emph{nondeterministic or} in \cite{Hoa85}.
Informally, a process $P \sqcap Q$ ``behaves like $P$ or like $Q$, where the 
selection between them is made arbitrary, without knowledge or control of
environment.''
In a formal operational semantics in the style of Plotkin [\citeyear{Plo80}],  
this is modeled by using different labels of transitions.
The special label $\tau$ appears at \emph{internal} or \emph{hidden} transitions, 
denoted by $p \stackrel{\tau}{\longrightarrow} q$, 
which cannot be controlled or even seen by the environment.
Labels $a \neq \tau$ appear at \emph{external} or \emph{visible} transitions,
denoted by $p \stackrel{a}{\longrightarrow} q$, and 
represent actions in which the environment can participate.
The selection of $P \sqcap Q$ is modeled by the internal transitions
$P \sqcap Q \stackrel{\tau}{\longrightarrow} P$ and 
$P \sqcap Q \stackrel{\tau}{\longrightarrow} Q$ \cite{Ros98,Ros10}.
Thus after this first hidden step, $P \sqcap Q$   behaves as $P$ or as~$Q$.

\emph{External nondeterminism} or \emph{alternation} is denoted by the
binary operator~$[]$, called \emph{general choice} in \cite{Hoa85}.
The idea of a process $P\, []\, Q$ is that ``the environment can
control which of $P$ and $Q$ will be selected, provided that this
control is exercised on the very first action.''  The formal
operational semantics of the operator~$[]$ in the style of Plotkin
\citeyear{Plo80} is more subtle than the one for $\sqcap$, see
\cite{OH86,Ros98,Ros10}.  
In applications,  $P\, []\, Q$ is performed in the context of 
a synchronous parallel composition 
with another process $R$ modeling a user or an environment.
Then the first visible transition with a label $a \neq \tau$ of $R$ 
has to synchronize with a first visible transition $P\, []\, Q$ 
with the same label $a$,
thereby selecting $P$ or $Q$ of the alternative $P\, []\, Q$. 
This formalizes Hoare's idea stated above.

These two nondeterministic operators have also been studied by
De Nicola and Hennessy in the context of Milner's CCS \cite{NicolaH87}. 
The authors write $\oplus$ for internal nondeterminism and
keep $[]$ for external nondeterminism. When defining the
operational semantics of the two operators, they
write $\longrightarrow$ instead of  $\stackrel{\tau}{\longrightarrow}$
and speak of ``CCS without $\tau$'s'' because $\tau$
is not present in this process algebra.
Subsequently, they introduce a testing semantics and provide
a complete algebraic characterization of the two operators.

To assess the effect of nondeterminism, the new CSP
introduced a new equivalence between processes due to
\cite{BHR84}, called \emph{failure equivalence}.  A \emph{failure} of
a process is a pair consisting of a trace, i.e., a finite sequence of
symbols that label transitions, and a set of symbols that after
performing the trace the process can \emph{refuse}.  Processes with
the same set of failures are called \emph{failure equivalent}.
Besides an equivalence, new CSP also provides a notion of \emph{refinement} among processes.
A process $P$ \emph{refines} a process $Q$ if the set of failures of $P$ is
a subset of the set of failures of $Q$.
Informally, this means that $P$ is \emph{more deterministic} than $Q$.
Thus by definition, processes that refine each other are failure equivalent.

As an example consider again the processes $P$ and $Q$ in Figure~\ref{fig:Milner}.
They are not failure equivalent, but $P$ refines $Q$.
This example shows that failure equivalence is sensitive to
nondeterminism. 
It turns out that failure equivalence coincides with the \emph{must} equivalence 
for `strongly convergent' processes, i.e., those without any
divergences \cite{DeNic87}. So it represents demonic nondeterminism.

\section{Conclusions}

As explained in this chapter, nondeterminism is a natural feature of
various formalisms used in computer science. The proposals put forward
prior to Dijkstra's seminal paper \cite{Dij75} are examples of what is
now called angelic nondeterminism.  Dijkstra's novel approach, now
called demonic nondeterminism, was clearly motivated by his prior work
on concurrent programs that are inherently nondeterministic in their
nature. His guarded command language became a simplest possible
setting allowing one to study demonic nondeterminism, unbounded
nondeterminism, and fairness.

Its versatility was demonstrated by subsequent works on diverse
topics.  In \cite{Mar86} correct delay-insensitive VLSI circuits were
derived by means of a series of semantics-preserving transformation
starting with a distributed programming language. In some aspects the
language is similar to CSP.  In its sequential part it uses a subset
of guarded commands with an appropriately customized semantics.  To
study randomized algorithms and their semantics an extension of the
guarded commands language with a probabilistic choice operator was
investigated in a number of papers, starting with \cite{JSM97}. More
recently, guarded commands emerged in the area of quantum programming,
as a basis for quantum programming languages, see, e.g., \cite{Yin16}.

As explained, the guarded commands language can also be viewed as a
stepping stone towards a study of concurrent programs.  In fact, it
can be seen as a logical layer that lies between deterministic and
concurrent programs.

The viability of Dijkstra's proposal can be best viewed by consulting
statistics provided by Google Scholar. They reveal that the original
paper, \cite{Dij75}, has been most often cited in the past decade.

\section*{Acknowledgements}

We would like to thank Nachum Dershowitz, Peter van Emde Boas, and
Reiner H\"{a}hnle for useful comments and Manfred Broy, Rob van
Glabbeek, and Leslie Lamport for suggesting a number of helpful
references.

% \backmatter

% \bibliography{acm-publications,nondet}
% \bibliographystyle{abbrv}
% \bibliographystyle{natbib}
 \bibliographystyle{abbrvnat}

\bibliography{ao23}

\end{document}